# Electronic Quantum Confinement in Cylindrical Potential Well

A. S. Baltenkov[1] and A. Z. Msezane[2]

[1]Institute of Ion-Plasma and Laser Technologies Tashkent 100125, Uzbekistan
[2]Center for Theoretical Studies of Physical Systems, Clark Atlanta University, Atlanta, Georgia 30314, USA

**Abstract.** The effects of quantum confinement on the momentum distribution of electrons confined within a cylindrical potential well have been analyzed. The motivation is to understand specific features of the momentum distribution of electrons when the electron behavior is completely controlled by the parameters of a non-isotropic potential cavity. It is shown that studying the solutions of the wave equation for an electron confined in a cylindrical potential well offers the possibility to analyze the confinement behavior of an electron executing one- or two-dimensional motion in the three-dimensional space within the framework of the same mathematical model. Some low-lying electronic states with different symmetries have been considered and the corresponding wave functions have been calculated; the behavior of their nodes and their peak positions with respect to the parameters of the cylindrical well has been analyzed. Additionally, the momentum distributions of electrons in these states have been calculated. The limiting cases of the ratio of the cylinder length $H$ and its radius $R_0$ have been considered; when the cylinder length $H$ significantly exceeds its radius $R_0$ and when the cylinder radius is much greater than its length. The cylindrical quantum confinement effects on the momentum distribution of electrons in these potential wells have been analyzed. The possible application of the results obtained here for the description of the general features in the behavior of electrons in nanowires with metallic type of conductivity (or nanotubes) and ultrathin epitaxial films (or graphene sheets) are discussed. Possible experiments are suggested where the quantum confinement can be manifested.

## 1. Introduction

The cylinder-like nanostructures, with the diameter of the order of a nanometer ($10^{-7}$cm) and the ratio of the length to width being greater than $10^2$-$10^3$, so called nanowires (nanowhiskers), have many interesting properties that are not seen in bulk or 3D (three-dimensional) materials and have vast potential for applications [1-9]. In nanowires with the metallic type of conductivity each of collectivized electrons in the first approximation can be considered to be locked in a thin and very long cylindrical potential box. The latter is created by the carbon ions, $C^{4+}(1s^2)$, (forming the positively charged cylindrical skeleton) and the collectivized $2s2p$-electrons themselves. At these scales of the potential box the quantum mechanical effects in electron behavior are important. Their behavior is defined by the solutions of the wave equation for a particle that is locked in the cylindrical potential well. Unlike the quantum confinement in the *spherically symmetric* potential boxes (quantum dots or carbon microspheres) in the given case we deal with quantum *non-isotropic confinement.* Variation of the cylindrical box parameters allows changes in the character of the motion of a particle in the box. In cases where the length of the cylinder significantly exceeds its radius we deal with almost the one-dimensional motion of the conduction electrons along the cylinder axis (1D electron gas). For simultaneously a small length and a large radius of the cylinder the electrons are approximately confined in a potential well having the form of a thin disk; the particle motion in this box is almost two-dimensional (2D electron gas). The special features of the cylindrical quantum confinement manifest themselves in the momentum distribution of the electrons as well. This paper is devoted to studying the momentum distribution of electrons confined within the cylindrical potential well.

The problem of measuring electron momentum distributions in atoms, molecules, and crystals is a long-standing one. The techniques used to obtain these data include Compton scattering of X rays [10-12], quasi-free electron scattering at large angles [13,14], positron annihilation [15,16], *etc*. The investigation of electron momentum distributions in oriented nanowires [17-19], ultrathin epitaxial films [20] or bunches of graphene sheets [21] by these methods expands significantly the existing ideas about the interaction of photons and charged particles with electronic structures of small 1D and 2D dimensions.

The special features of the electron momentum distribution in anisotropic systems can be understood qualitatively from the consideration of a simple model system – electrons bound by a linear chain of the atomic potentials (Sec. 1). The wave functions and eigenenergy values of electrons confined within the cylindrical potential well are calculated in Section 2. These wave functions are used in Section 3 to calculate the electron momentum distribution in these stationary states. The limiting cases of the ratio of the cylinder length $H$ and its radius $R_0$ are considered in Section 4; firstly, when the cylinder length $H$ significantly exceeds its radius $R_0$ and then when the cylinder radius is much greater than its length. The numerical results for the electron momentum distributions for some low-lying states in the cylindrical potential well are presented in Section 5. Section 6 presents the Discussions and the Conclusions.

## 2. Electron in a linear chain of atomic potentials

Let us consider the behavior of an electron bound by a linear chain of atomic potentials. This system of potentials is strongly anisotropic. The electron is delocalized along the chain axis Z and a degree of delocalization is defined by the interatomic distance $R$. At the same time the electron is strongly localized in the directions perpendicular to the Z axis. The ground (symmetric) bound state of electrons in the linear chain of $N$ atomic potentials ($N$>>1) can be written in the zero approximation as a linear combination of the atomic functions

$$\psi(\mathbf{r}) = C\sum_{n=0}^{N-1}\varphi(\mathbf{r}+n\mathbf{R})\,. \tag{1}$$

Here $\mathbf{R}=\mathbf{v}R$ where $\mathbf{v}$ is the unit vector along the chain Z-axis. We will consider that the distance between the potential wells $R$ is greater than the size of the electron localization region near each of the atomic nuclei. In this case the binding energy of an electron in the linear chain is close to that of an isolated electron in a single atom. Then the overlap integral is equal to zero and the normalization factor is $C=N^{-1/2}$. The wave function (1) with $N$=2 was used by Cohen and Fano [22] to study the interference of molecule photoionization.

The electron momentum distribution is defined by the Fourier transform of the wave function (1)

$$\psi(\mathbf{k}) = C\sum_{n=0}^{N-1}\int\varphi(\mathbf{r}+n\mathbf{R})e^{-i\mathbf{k}\cdot\mathbf{r}}d\mathbf{r} = C\varphi(k)\sum_{n=0}^{N-1}e^{in\mathbf{k}\cdot\mathbf{R}}\,. \tag{2}$$

Here $\varphi(k)$ is the Fourier-image of the isolated atom wave function

$$\varphi(k) = \int e^{-i\mathbf{k}\cdot\mathbf{r}}\varphi(\mathbf{r})d\mathbf{r}\,. \tag{3}$$

The sum of the geometric progression in (2) is determined by the following expression

$$\sum_{n=0}^{N-1} e^{in\mathbf{k}\cdot\mathbf{R}} = \frac{1-e^{iN\mathbf{k}\cdot\mathbf{R}}}{1-e^{i\mathbf{k}\cdot\mathbf{R}}} . \quad (4)$$

The modulus squared of the wave function (2) is the probability distribution of the different values of momentum for an electron bound by the linear potential chain

$$|\psi(\mathbf{k})|^2 \propto \frac{\sin^2[N(\mathbf{k}\cdot\mathbf{R})/2]}{\sin^2[(\mathbf{k}\cdot\mathbf{R})/2]} . \quad (5)$$

The extreme strong dependence of the electron angular distribution (5) on the angle between the momentum vector **k** and the potential chain axis **R** is evident. In particular, when the electron momentum vector is perpendicular to the chain axis the scalar product $(\mathbf{k}\cdot\mathbf{R}) \to 0$, so the ratio in (5) goes to $N^2 >> 1$.

Thus, the momentum distribution of electrons bound by a linear chain of the atomic potentials (5) has sharp giant maximum in the plane perpendicular to the chain axis*. The similar picture we can expect for electron locked in a long cylindrical potential well with the radius much less than its length. In this trap the electron is also delocalized along the cylinder axis but strongly localized in the perpendicular directions. The difference is that in the cylindrical potential well the electron freely moves along the cylinder axis while in the atomic chain it moves along the Z-axis in the periodical potential.

### 3. Wave equation

Consider the motion of single electron in the circular cylindrical potential well formed by the positively charged cylindrical core (formed by the smeared $C^{4+}$ ions) and negatively charged cloud of collectivized 2*s*2*p*-electrons. Here we use the approach that is widely used for spherically symmetric fullerenes (see for example [24]). Namely, we replace the electric charge of nanowire skeleton by the uniform distribution of the positive charge (distribution having the cylinder-like form) in the field of which the valence electrons move.

Let us suppose that $R_0$ and $H$ are the radius and height of the cylindrical potential box, respectively. In the cylindrical coordinate system with the Z-axis coinciding with that of the cylinder the wave equation in atomic units ($|e| = m = \hbar = 1$) has the form:

$$-\frac{1}{2}\left[\frac{1}{\rho}\frac{\partial}{\partial\rho}\left(\rho\frac{\partial}{\partial\rho}\right) + \frac{1}{\rho^2}\frac{\partial^2}{\partial\varphi^2} + \frac{\partial^2}{\partial z^2}\right]\psi = E\psi . \quad (6)$$

Here we consider the cylinder walls to be impenetrable for a particle so that the wave function is $\psi = 0$ on the surface of the cylindrical potential well. The wave function solution is then of the form

$$\psi = R(\rho)Z(z)\Phi(\varphi) . \quad (7)$$

After separation of variables we obtain the following equation for the function $Z(z)$

$$\frac{\partial^2 Z}{\partial z^2} = -2E_z Z = -k_z^2 Z . \quad (8)$$

Since the cylinder walls are impenetrable for a particle, the boundary conditions for the function Z($z$) are of the form

$$Z(0) = Z(H) = 0 \quad (9)$$

The solution of Eq. (8) is written as

---

* Besides this giant maximum, the function (5) has also the maximums of much less intensity. See Fraunhofer diffraction by a grating with $N$ slits [23].

$Z(z) = a\sin(k_z z + \delta)$

Hence, the following expression is obtained for the function *Z*(*z*)

$$Z_n(z) = a\sin(\frac{\pi n}{H}z) \tag{10}$$

Here *n* corresponds to the positive integers beginning with one and the wave vector $k_z = \pi n / H$. For the function $\Phi(\varphi)$ we have the expression

$$\frac{\partial^2 \Phi}{\partial \varphi^2} = -m^2 \Phi \tag{11}$$

The solutions of this equation are $\Phi = \exp(im\varphi)$ with the condition that *m* is a positive integer beginning with zero. Therefore, we obtain the following equation for the function $R(\rho)$

$$\rho^2 \left( \frac{\partial^2 R}{\partial \rho^2} \right) + \rho \left( \frac{\partial R}{\partial \rho} \right) + (\rho^2 k^2 - m^2) R = 0, \tag{12}$$

where $k^2 = 2(E - E_z)$. The finite solution of this equation is the Bessel function [25]

$$R_m(\rho) = J_m(k\rho). \tag{13}$$

The boundary condition for the wave function $R(\rho)$ leads to the following equation

$$R_m(R_0) = J_m(kR_0) = 0. \tag{14}$$

Let $x = q_{ml}$ be the *l*-root of the Bessel function $J_m(x)$. From Eq. (14) we obtain the following quantization condition for the wave vector *k*:

$$k_{ml} = q_{lm} / R_0. \tag{15}$$

Corresponding to this wave vector, the function $R(\rho)$ is now determined by the two quantum numbers: the root number *l* of the Bessel function and its order *m* and has the form $R_{lm}(\rho) = J_m(q_{lm}\rho / R_0)$.

The total electron energy in the cylindrical potential well is determined by the quantum numbers *n*, *l* and *m*. For this energy (in the usual units) we have the following expression [26]

$$E_{nlm} = \frac{k_{ml}^2}{2} + \frac{k_z^2}{2} = \frac{\hbar^2}{2m}\left[ \left( \frac{q_{lm}}{R_0} \right)^2 + \left( \frac{n\pi}{H} \right)^2 \right]. \tag{16}$$

The electron wave function corresponding to this energy is given by the following expression

$$\psi_{nlm}(\rho, z, \varphi) = N_{nlm} J_m \left( \frac{q_{lm}}{R_0} \rho \right) \sin\left( \frac{\pi n}{H} z \right) e^{im\varphi}. \tag{17}$$

The quantum numbers *n* and *l* here are the positive integers beginning with one, *m*=0,1… The normalization factor $N_{nlm}$ is defined as usual by

$$N_{nlm}^2 2\pi \int_0^{R_0} J_m^2 \left( \frac{q_{lm}}{R_0} \rho \right) \rho d\rho \int_0^H \sin^2 \left( \frac{\pi n}{H} z \right) dz = 1. \tag{18}$$

The integrals in this expression are

$$\int_0^H \sin^2 \left( \frac{\pi n}{H} z \right) dz = \frac{H}{2};$$

$$\int_0^{R_0} J_m^2 \left( \frac{q_{lm}}{R_0} \rho \right) \rho d\rho = \frac{1}{2} R_0^2 J_{m+1}^2 (q_{lm}). \tag{19}$$

Then the normalization factor becomes

$$N_{nlm}^2 = \frac{2}{\pi H R_0^2 J_{m+1}^2(q_{lm})} = \left[\frac{1}{2\pi}\right]\left[\frac{2}{H}\right]\left[\frac{2}{R_0^2 J_{m+1}^2(q_{lm})}\right]. \quad (20)$$

The electron ground state in the cylindrical potential well is described by the wave function

$$\psi_{110}(\rho, z, \varphi) = N_{110} J_0\left(\frac{q_{10}}{R_0}\rho\right)\sin\left(\frac{\pi}{H}z\right); \quad (21)$$

where $q_{10} \approx 2.4048$ is the value of the first root of the Bessel function $J_0(x)$ [25].

The behavior of the function $Z_1(z)$ in (21) that describes the one-dimensional free motion of electron along the cylinder axis is evident. The functions $R_{lm}(\rho)$ for several quantum numbers $l$ and $m$ are presented in Fig. 1. The cylinder radius $R_0$ in these calculations is equal to unity. Some roots of the Bessel functions $q_{ml}$ and the values of the functions $J_{m+1}(q_{lm})$ used in these calculations are presented in Table 1.

Table 1

| $lm$ | $q_{lm}$ | $J_{m+1}(q_{lm})$ |
|---|---|---|
| 10 | 2.4048 | 0.5195 |
| 20 | 5.5201 | -0.3403 |
| 30 | 8.6537 | 0.2715 |
| 11 | 3.8317 | 0.4024 |
| 21 | 7.0156 | -0.3005 |
| 31 | 10.1735 | 0.2493 |

**4. The Fourier transform of wave function**

The bound state wave function $\psi_{nlm}$ (17) in the momentum representation is defined by the following integral

$$\psi_{nlm}(\mathbf{k}) = \int \psi_{nlm}(\mathbf{r})\exp[i(\mathbf{k}\cdot\mathbf{r})]d\mathbf{r} = \int \psi_{nlm}(\rho, z, \varphi)\exp[i(k_x\rho\cos\varphi + k_z z)]\rho d\rho dz d\varphi. \quad (22)$$

Here $k_x, k_z$ are the Cartesian components of electron momentum **k**. Because of the cylindrical symmetry of the problem the $y$-component of vector **k** can be set equal to zero. In view of Eq. (17) we write the integral (22) as

$$\psi_{nlm}(k_x, k_z) = N_{nlm}\int_0^{R_0} J_m\left(\frac{q_{lm}}{R_0}\rho\right)\rho d\rho \int_0^H \sin\left(\frac{\pi n}{H}z\right)e^{ik_z z}dz \int_0^{2\pi} e^{ik_x\rho\cos\varphi + im\varphi}d\varphi. \quad (23)$$

The integrals in this expression are

$$I_\varphi = \int_0^{2\pi} e^{i(k_x\rho\cos\varphi + m\varphi)}d\varphi = 2\pi i^m J_m(k_x\rho);$$

$$I_z = \int_0^H e^{ik_z z}\sin\left(\frac{\pi n}{H}z\right)dz = \frac{\pi n H}{(\pi n)^2 - (Hk_z)^2}[1 - e^{i(\pi n + k_z H)}];$$

$$I_\rho = 2\pi i^m \int_0^{R_0} J_m(k_x\rho)J_m\left(\frac{q_{lm}}{R_0}\rho\right)\rho d\rho = \frac{2\pi i^m R_0}{k_x^2 - (q_{lm}/R_0)^2}\frac{q_{lm}}{R_0}J_m(k_x R_0)J'_m(q_{lm}). \quad (24)$$

Taking into account the formulas (24), we write the function $\psi_{nlm}(k_x, k_z)$ as:

$$\psi_{nlm}(k_x,k_z)=N_{nlm}I_\rho I_z\ . \tag{25}$$

The electron momentum distribution $\Phi(k,\vartheta_k)$ is proportional to the modulus squared of the bound state wave function in the momentum representation

$$\Phi(k,\vartheta_k)\propto|\psi_{nlm}(k_x,k_z)|^2\propto\left[\frac{J_m(k_xR_0)}{q_{lm}^2-(k_xR_0)^2}\right]^2\left[\frac{\sin[(\pi n+k_zH)/2]}{(\pi n)^2-(k_zH)^2}\right]^2 . \tag{26}$$

Here the spherical components of the electron momentum are $k_x=k\sin\vartheta_k$ and $k_z=k\cos\vartheta_k$. Since we are interested in the shape of the momentum distribution we omitted some insignificant constants in the expression (26). The denominators and numerators of the two fractions in (26) go to zero for $k_xR_0\to q_{lm}$ and $k_zH\to n\pi$. Using the L'Hôpital's rule we obtain the following values of the fractions within these limits

$$\lim\frac{J_m(k_xR_0)}{q_{lm}^2-(k_xR_0)^2}=-\frac{J'_m(q_{lm})}{2q_{lm}};$$

$$\lim\frac{\sin[(\pi n+k_zH)/2]}{(\pi n)^2-(k_zH)^2}=\frac{(-1)^{n+1}}{4\pi n}. \tag{27}$$

Further, we use the general expression (26) to calculate the electron momentum distribution for some stationary states.

## 5. Numerical calculations

### 5.1 Case *H*>>*R*₀

For this condition on the potential well parameters the quantization of electron energy is defined by Eq. (16), having the following form

$$E_{nlm}\approx\frac{\hbar^2}{2m}\left(\frac{q_{lm}}{R_0}\right)^2 . \tag{28}$$

Thus, the energy levels in the potential well are determined by a pair of the quantum numbers *lm* only, i.e. the number of the roots of the Bessel function *l* and its index *m*; they also define a characteristic wave vector of the electron $k_{lm}=q_{lm}/R_0$ in this stationary state. The function (26) is a product of two functions; the first one $F_1(k,\vartheta_k)$ depends on the quantum numbers *lm*, while the second $F_2(k,\vartheta_k)$ is a function of the quantum number *n*. For fixed momentum *k* they, as functions of the polar angle $\theta_k$ (within the range of angles $0\le\vartheta_k\le\pi$), are characterized by principally different behavior. On the edges of this angular range the first function $F_1(k,\vartheta_k)$ is maximal for *m*=0 and equal to zero for other indices *m* because of the vanishing of the Bessel function $J_{m\neq0}(0)=0$. The second function $F_2(k,\vartheta_k)$ has a value of order $(kH)^{-4}$ within almost the whole range of angles except those where the denominator of the second fraction is close to zero. At those points the function $F_2(k,\vartheta_k)$ has the maximal value of $(4\pi n)^{-2}$. Therefore, the electron angular momentum $\theta_k$-distribution $\Phi(k,\vartheta_k)$ for *H*>>*R*₀ is fully determined by the behavior of the function $F_2(k,\vartheta_k)$ having the form

$$F_2(k,\vartheta_k)=\left[\frac{\sin[(\pi n+k_zH)/2]}{(\pi n)^2-(k_zH)^2}\right]^2=\frac{1}{2[(\pi n)^2-(kH\cos\vartheta_k)^2]^2}\begin{cases}[1-\cos(kH\cos\vartheta_k)], & n \text{ is even;}\\ [1+\cos(kH\cos\vartheta_k)], & n\text{ is odd.}\end{cases} \tag{29}$$

The function (29) together with the function $\Phi(k,\vartheta_k)$ for even quantum numbers *n* goes to zero for $\theta_k=\pi/2$; for the odd ones these functions are different from zero. The denominator of the second fraction in (29) vanishes for angles $\vartheta_k^{res} = \arccos(n\pi / kH)$ and $\vartheta_k^{res} = \pi - \arccos(n\pi / kH)$, leading to the appearance of two main resonance peaks. For $kH \gg n\pi$ these peaks tend to merge into one resonance peak at the point $\theta_k = \pi/2$.

The electron angular momentum distributions $\Phi(k, \vartheta_k)$ calculated with the formula (26) for selected low-lying electronic states $\psi_{nlm}$ are presented in Fig. 2. In these calculations the potential well parameters typical for nanowires $R_0$= 20 au≈$10^{-7}$ *cm*, *H*=2000 au≈$10^{-5}$ *cm* were used. The left panel corresponds to the odd numbers *n*=1-5; the right one to the even numbers *n*=2-6. All the curves in this figure are normalized to unity at the functions maxima. In each of the graphs the two curves $\Phi(k, \vartheta_k)$ and $F_2(k, \vartheta_k)$ are presented. Coincidence of these curves is evidence that the variations of the function $F_1(k, \vartheta_k)$ in the vicinity of the point $\theta_k \approx \pi/2$ have no influence on the shape of the electron momentum distribution, i.e. for this potential well geometry ($H \gg R_0$) the normalized function $\Phi(k, \vartheta_k)$ is independent of the quantum numbers *lm*. In the calculations they were assumed to be equal to *l*=1 and *m*=0; the wave vector *k* was fixed at $k_{10} = q_{10}/R_0 \approx 0.1202$ au.

The electron angular momentum distributions connected with the linear chain of the potentials (5) and the electron in the ground state (21) in the cylindrical potential well are presented in polar coordinates in Fig. 3. The length of the vector **k** in this figure is proportional to the probability of existence of the momentum **k** for the given angular momentum distribution; $\theta_k$ is the angle between the vectors **k** and the polar Z-axis. With the increase in the number of atoms in the chain *N* or in the ratio of the cylinder length to its radius $H/R_0$, the curves in this figure degenerate into a line perpendicular to the Z-axis. The 3D pictures of the electron momentum $\theta_k$-distribution are the figures of rotation of these curves around the Z-axis being the axis of the cylinder of the atomic chain. Thus, the electron momentum vectors **k** are oriented mainly perpendicular to the cylinder axis.

For the fixed angle between the vectors **k** and Z ($\theta_k \approx \pi/2$) the function (26), as a function of *k*, describes the distribution of lengths of electron momentum vectors **k** in the XY-plane and it coincides within a constant with the function $F_1(k, \vartheta_k)$. For the lowest electronic states $\psi_{nlm}$ the electron momentum *k*-distributions $k^2 F_1(k, \pi/2)$, normalized by the condition

$$\int_0^\infty F_1(k, \pi/2) k^2 dk = 1, \tag{30}$$

are presented in Fig. 4. In each of the graphs in Fig. 4 the arrow points to the momenta $k_{lm} = q_{lm}/R_0$, characteristic of this *lm* state. Thus, according to Figs. 2-4, for $H \gg R_0$ the electron momentum vectors **k** are mainly in the XY-plane perpendicular to the cylinder axis and their lengths are concentrated mainly in the vicinity of the momenta $k_{lm}$.

### 5.2 Case $R_0 \gg H$

For this ratio of the potential well parameters the quantization of the electron energy is defined by the equation (16) having the form

$$E_{nlm} \approx \frac{\hbar^2}{2m} \left( \frac{n\pi}{H} \right)^2 . \tag{31}$$

Thus, the energy levels in the potential well are determined by the quantum number *n*; it also defines the characteristic wave vector of the electron $k_n = \pi n / H$ in the given stationary state. The electron momentum vectors **k** for $R_0 \gg H$ are concentrated mainly within the narrow cone of angles around the Z-axis, i.e. at the angles $\theta_k \approx 0$ and $\theta_k \approx \pi$. In the vicinity of these points the electron angular momentum $\theta_k$-distribution $\Phi(k,\vartheta_k)$ is fully determined by the behavior of the function $F_1(k,\vartheta_k)$ having the form

$$F_1(k,\vartheta_k)=\left[\frac{J_m(kR_0\sin\vartheta_k)}{q_{lm}^2-(kR_0\sin\vartheta_k)^2}\right]^2. \qquad (32)$$

The variation of the function $F_2(k,\vartheta_k)$ in the vicinity of these points has no effect on the shape of the electron momentum distribution, i.e. for this potential well geometry ($R_0$>>$H$) the normalized function $\Phi(k,\vartheta_k)$ is independent of the quantum number $n$. The function (32) together with the function $\Phi(k,\vartheta_k)$ for $\theta_k$=0 and $\theta_k\approx\pi$ are different from zero for $m$=0 only; for all other quantum numbers $m$ the functions (26) and (32) vanish at these points. The denominator of the fraction in (32) goes to zero for angles $\vartheta_k^{res}=\arcsin(q_{lm}/kR_0)$ and $\vartheta_k^{res}=\pi-\arcsin(q_{lm}/kR_0)$, leading to the appearance of the main resonance peaks in the electron angular momentum distributions. For $kR_0$>>$q_{lm}$ these peaks tend to merge into one resonance peak at the points $\theta_k$=0 and $\theta_k\approx\pi$.

The electron angular momentum distributions $\Phi(k,\vartheta_k)$ calculated within the angular range $-\pi/2<\theta_k<\pi/2$ with the formula (32) for some low-lying electronic states $\psi_{nlm}$ are presented in Fig. 5. In these calculations the parameters of the potential well $H$=20 au≈$10^{-7}$ *cm* and $R_0$=2000 au≈$10^{-5}$ *cm* were used. The left panel corresponds to $m$=0 and $l$=1-3; the right one to $m$=1 and $l$=1-3. All the curves in this figure are normalized to unity at the maxima of the functions. In these calculations of the $\theta_k$-distribution of electron momentum the quantum number $n$ was equal to unity, $n$=1; the wave vector $k$ was fixed and equal to $k_1=\pi/H\approx0.1571$ au.

Fig. 6 presents in polar coordinates the $\theta_k$-distribution of electron momentum in the ground $\psi_{110}$ state. The vector length $k$ in this figure (as in Fig. 3) is proportional to the probability of existence of momentum **k** for the given angular momentum distribution; $\theta_k$ is the angle between the vectors **k** and polar Z-axis. With the increase in the ratio of the cylinder radius to its length, namely $R_0/H$>>1 the curves in this figure degenerate into a segment along the Z-axis. The 3D pictures of the electron momentum $\theta_k$-distribution represent the figures of rotation of these curves around the Z-axis, also being the cylinder axis.

For the zero angle between the vectors **k** and **Z** the function (26), as the function $k$, describes the distribution of lengths of electron momentum vectors $k$ along the Z-axis and this function coincides within a constant with the function $F_2(k,\vartheta_k)$. The normalization condition of this function is defined similarly to (30)

$$\int_0^\infty F_2(k,0)k^2dk=1. \qquad (33)$$

The electron momentum $k$-distributions $k^2F_2(k,0)$ for the six lowest electronic states $\psi_{nlm}$ are presented in Fig. 7. Their maxima are close to the values of the wave vectors $k_n=\pi n/H$ typical for these states. Thus, the electron momentum vectors **k** in the case $R_0$>>$H$ are oriented mainly along the cylinder axis and the lengths of these vectors have a value close to $k_n$.

## 6. Discussions and Conclusions

The above-described specific features of the momentum distribution of electrons subjected to cylindrical confinement, namely a pronounced asymmetry of the momentum distribution of bound electrons, manifest themselves in diverse elementary processes. Under specific conditions this anisotropy can be observed during the process of target photoionization. Let us represent the wave function of an electron ionized out of a target as a plane wave (as in [22]). A dipole matrix element of the electron transition from the electron bound state to the continuum is defined as follows

$$M = -i\int e^{-i\mathbf{k}\cdot\mathbf{r}}(\mathbf{e}\cdot\nabla)\psi(\mathbf{r})d\mathbf{r}\ . \tag{34}$$

Integrating (34) by parts, we obtain the following expression for the matrix element

$$M = (\mathbf{e}\cdot\mathbf{k})\int \psi(\mathbf{r})e^{-i\mathbf{k}\cdot\mathbf{r}}d\mathbf{r} = (\mathbf{e}\cdot\mathbf{k})\psi(\mathbf{k})\ . \tag{35}$$

The differential cross section for photoionization $d\sigma / d\Omega$ is proportional to the absolute square of this matrix element:

$$\frac{d\sigma}{d\Omega} \propto (\mathbf{e}\cdot\mathbf{k})^2 \,|\,\psi(\mathbf{k})\,|^2\ . \tag{36}$$

Hence, the shape of the angular distribution of photoelectrons knocked out of the target in the Born approximation when the photon energy is much higher than the potential of target ionization is determined by the momentum distribution of electrons in the bound state. Substituting $|\,\psi(\mathbf{k})\,|^2$ from (5) with $N$=2 to (36), we obtain the formula (4) in [27] where the angular distribution of photoelectrons knocked out from the fixed-in-space two homo nuclear molecules was considered.

The anisotropy of the momentum distribution of electrons can be found in the ($e$, 2$e$) experiments. The phenomenon of impact ionization of a target by fast electrons is the basis of electron momentum spectroscopy [28-30]. The essence of this method is as follows. A beam of fast electrons is incident upon a target. With the help of a coincidence scheme, one selects events from the huge number of events caused by the beam of electrons in which an incident electron knocks out a target electron by means of the Coulomb interaction, transferring to it a significant part of the incident electron's kinetic energy. For this process both the energy and angular distribution of the final electron are measured. The differential cross section for impact ionization in the Plane-Wave Impulse Approximation (PWIA) is proportional to the absolute square of the bound-state wave function in the momentum representation:

$$\frac{d^3\sigma}{dE_i d\Omega_i d\Omega_e} \propto |\,\psi(\mathbf{k})\,|^2\ . \tag{37}$$

Here $d\Omega_i$ and $d\Omega_e$ are, respectively the elements of solid angles of emission of the incident and ejected fast electrons, and $dE_i$ is the energy spread of the knocked-out electron.

The photon angular distribution for two-quantum annihilation of positrons with electrons in media is also defined by the momentum distribution of pair annihilation [15, 16, 31]. So, the anisotropy of the momentum distribution of electrons in oriented graphite that can be considered as a bunch of graphene sheets [21] (in each of them the 2D motion of electrons is realized) was observed in experiments [32]. One can expect the similar effects in the ultrathin epitaxial graphite films [20]. These objects also can serve as a model of the molecular structures with 2D electron gas behavior.

Experiments show that under certain conditions on the substrate surface the matrices composed of vertically oriented nanowires [17-19] or nanotubes [33-35] can be formed. Such matrices can serve as objects for the experimental study of structures where the almost 1D motion of collectivized electrons is realized.

Finally, we note that although we have studied some low-lying electronic states in the cylindrical potential well, it is quite evident that the general special features of electron momentum distribution considered here will be also inherent to any states of electron being under the cylindrical quantum confinement conditions. We hope that the study performed in this paper would be applicable to the interpretation of the above-referenced experiments.

**Acknowledgments**

This work was supported by the Uzbek Foundation Award Ф2-ФА-Ф164 (ASB) and U.S. DOE, Basic Energy Sciences, Office of Energy Research (AZM).

**References**

1. H. Sakaki, Jpn. J. Appl. Phys., **19** L735 (1980).
2. R. S. Wagner, W. C. Ellis, Appl. Phys. Lett., **4** 89 (1964).
3. G. Zheng, W. Lu, S. Jin, C.M. Lieber, Adv. Mater., **16** 1890 (2004).
4. A. B. Greytak, L. J. Lauhon, M. S. Gudiksen, C. M. Lieber. Appl. Phys. Lett., **84** 4176 (2004).
5. Y. Li, J. Xiang, F. Qian, S. Gradečak, Y. Wu, H. Yan, D. A. Blom, C. M. Lieber, Nano Lett., **6** 1468 (2006).
6. S. Gradečak, F. Qian, Y. Li, H.-G. Park, C. M. Lieber, Appl. Phys. Lett., **87** 173111 (2005).
7. F. Patolsky, G. Zheng, O. Hayden, M. Lakadamyali, X. Zhuang, C. M. Lieber, Proc. Natl. Acad. Sci. USA, **101** 14017 (2004).
8. H. A. Nilsson, C. Thelander, L. E. Fröberg, J. B. Wagner, L. Samuelson, Appl. Phys. Lett., **89** 163101 (2006).
9. K. A. Dick, K. Deppert, T. Mårtensson, S. Mandl, L. Samuelson, W. Seifert, Nano Lett., **5** 761 (2005).
10. C. Utfeld, J. Laverock, T. D. Haynes, S. B. Dugdale, J. A. Duffy, M. W. Butchers, J. W. Taylor, S. R. Giblin, J. G. Analytis, J.-H. Chu, I. R. Fisher, M. Itou, Y. Sakurai, Phys. Rev. B **81** 064509 (2010).
11. J. Laverock, S. B. Dugdale, J. A. Duffy, J. Wooldridge, G. Balakrishnan, M. R. Lees, G.-q. Zheng, D. Chen, C. T. Lin, A. Andrejczuk, M. Itou, Y. Sakurai, Phys. Rev. B **76** 052509 (2007).
12. S. B. Dugdale, R. J. Watts, J. Laverock, Zs. Major, M. A. Alam, M. Samsel-Czekała, G. Kontrym-Sznajd, Y. Sakurai, M. Itou, D. Fort, Phys. Rev. Lett. **96** 046406 (2006)
13. A. L. Hughes, M. M. Mann, Phys. Rev. **53** 50 (1938).
14. R. Camilloni, A. Giardini Guidoni, R. Tiribelli, G. Stefani, Phys. Rev. Lett. **29** 618 (1972).
15. P. E. Mijnarends, Phys. Rev. **160** 512 (1967), and references cited therein.
16. S.B. Dugdale, J. Laverock, C. Utfeld, M. A. Alam, T. D. Haynes, D. Billington, D. Ernsting, Journal of Physics: Conference Series **443** 012083 (2013).
17. J. Noborisaka, J. Motohisa, T. Fukui, Appl. Phys. Lett. **86** 213102 (2005).
18. P. Mohan, J. Motohisa, T. Fukui, Nanotechnology **16** 2903 (2005).
19. J. Noborisaka, J. Motohisa, S. Hara, T. Fukui, Appl. Phys. Lett. **87** 093109 (2005).
20. C. Berger, Z. Song, T. Li, X. Li, A. Y. Ogbazghi, R. Feng, Z. Dai, A. N. Marchenkov, E. H. Conrad, P. N. First, W. A. de Heer, J. Phys. Chem. B **108** 19912 (2004).
21. X. Huang, H. Zhang, National Science Review **2** 19 (2015).
22. H. D. Cohen, U. Fano, Phys. Rev. **150** 30 (1966).
23. L. D. Landau and E. M. Lifshitz, *Theory of Fields*. Pergamon Press, Oxford, 1968.
24. A. V. Verkhovtsev, R. G. Polozkov, V. K. Ivanov, A. V. Korol and A. V. Solov'yov, J. Phys. B: At. Mol. Opt. Phys. **45** 215101 (2012).
25. M. Abramowitz, I. A. Stegun, *Handbook of Mathematical Functions*, National Bureau of Standards, Applied Mathematics Series **55** (1964).
26. G. Arfken *Mathematical Methods for Physicists*, Academic Press New York and London, (1970).
27. A. S. Baltenkov, U. Becker, S. T. Manson and A. Z. Msezane, J. Phys. B: At. Mol. Opt. Phys. **45** 035202 (2012).
28. R. Camilloni, A. Giardini Guidoni, R. Tiribelli, G. Stefani, Phys. Rev. Lett. **29** 618 (1972).
29. E. Weigold, S. T. Hood, P. J. O. Teubner, Phys. Rev. Lett. **30** 475 (1973).
30. S. T. Hood, I. E. McCarthy, P. J. O. Teubner, E. Weigold, Phys. Rev. A **8** 2494 (1973).

31. S. Berko, F. L. Hereford, Rev. Mod. Phys. **28** 299 (1956).
32. S. Berko, R. E. Kelley, J. S. Plaskett, Phys. Rev. **106** 824 (1957).
33. S. Fan, W. Liang, H. Dang, N. Franklin, T. Tombler, M. Chapline, H. Dai, Physica E **8** 179 (2000).
34. W. A. De Heer, A. Châtelain, D. Ugarte, Science **270** 1179 (1995).
35. Y. Chen, D. T. Shaw, L. Guo, Appl. Phys. Lett. **76** 2469 (2000).

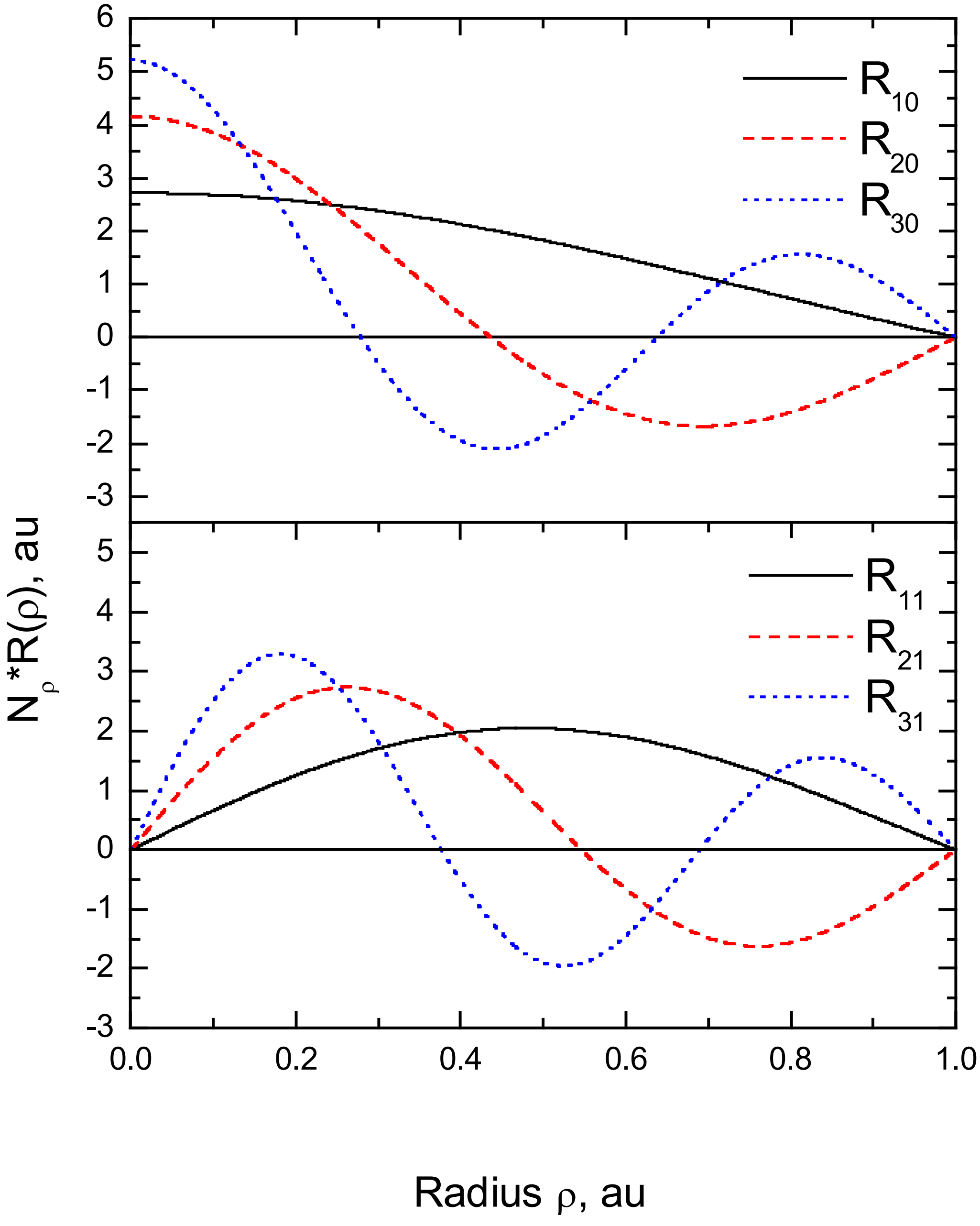

Fig. 1. The normalized functions $R_{lm}(\rho)$ for some quantum numbers $l$ and $m$. The radius of the cylindrical potential well is $R_0$=1 au.

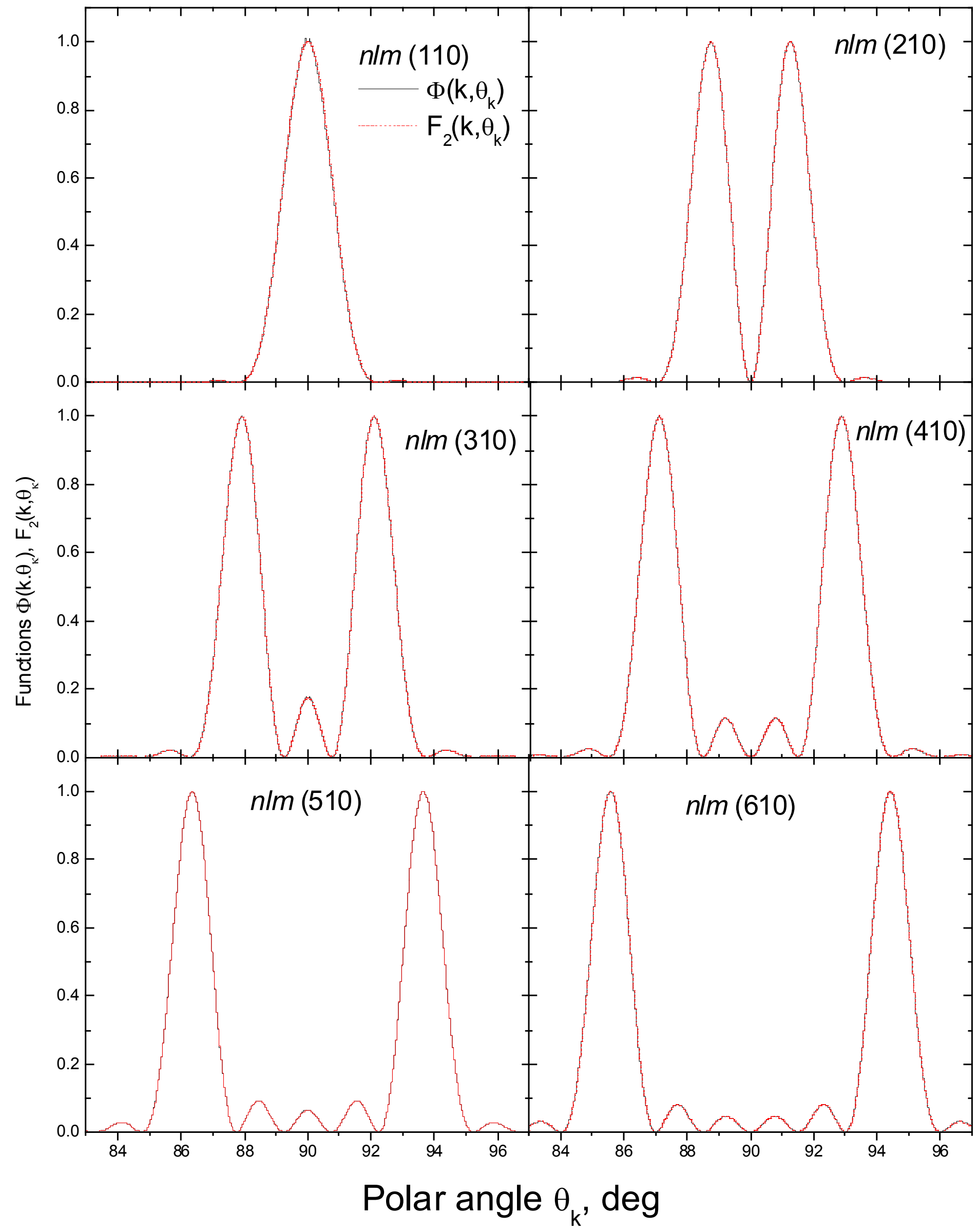

Fig. 2. The angular distributions of electron momentum $\Phi(k,\vartheta_k)$ and functions $F_2(k,\vartheta_k)$ for some low-lying electronic states $\psi_{nlm}$

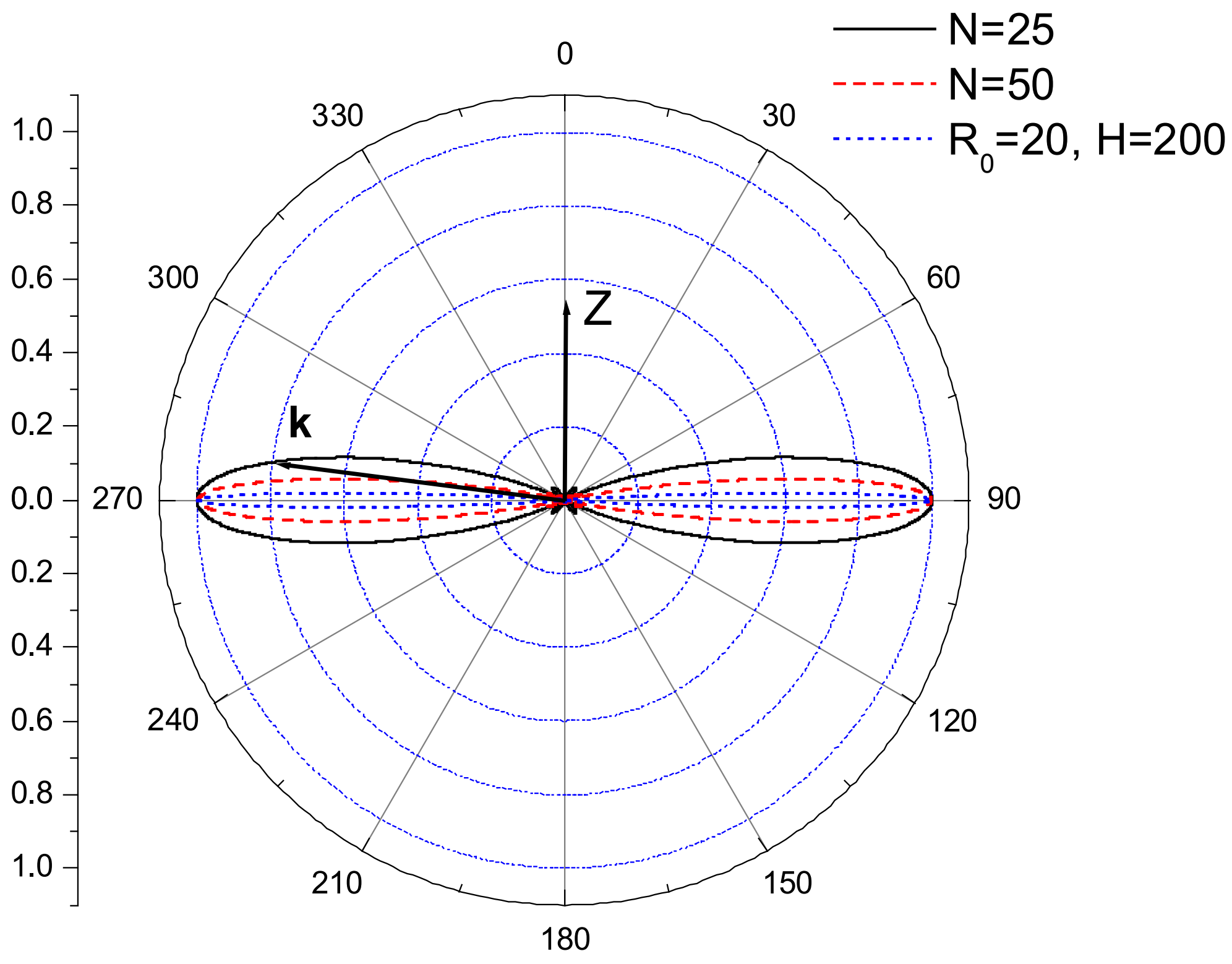

Fig. 3. The electron angular momentum distribution in the linear atomic chain (5) and of electron in the ground state (21) confined in the "long" cylindrical potential well (polar coordinates)

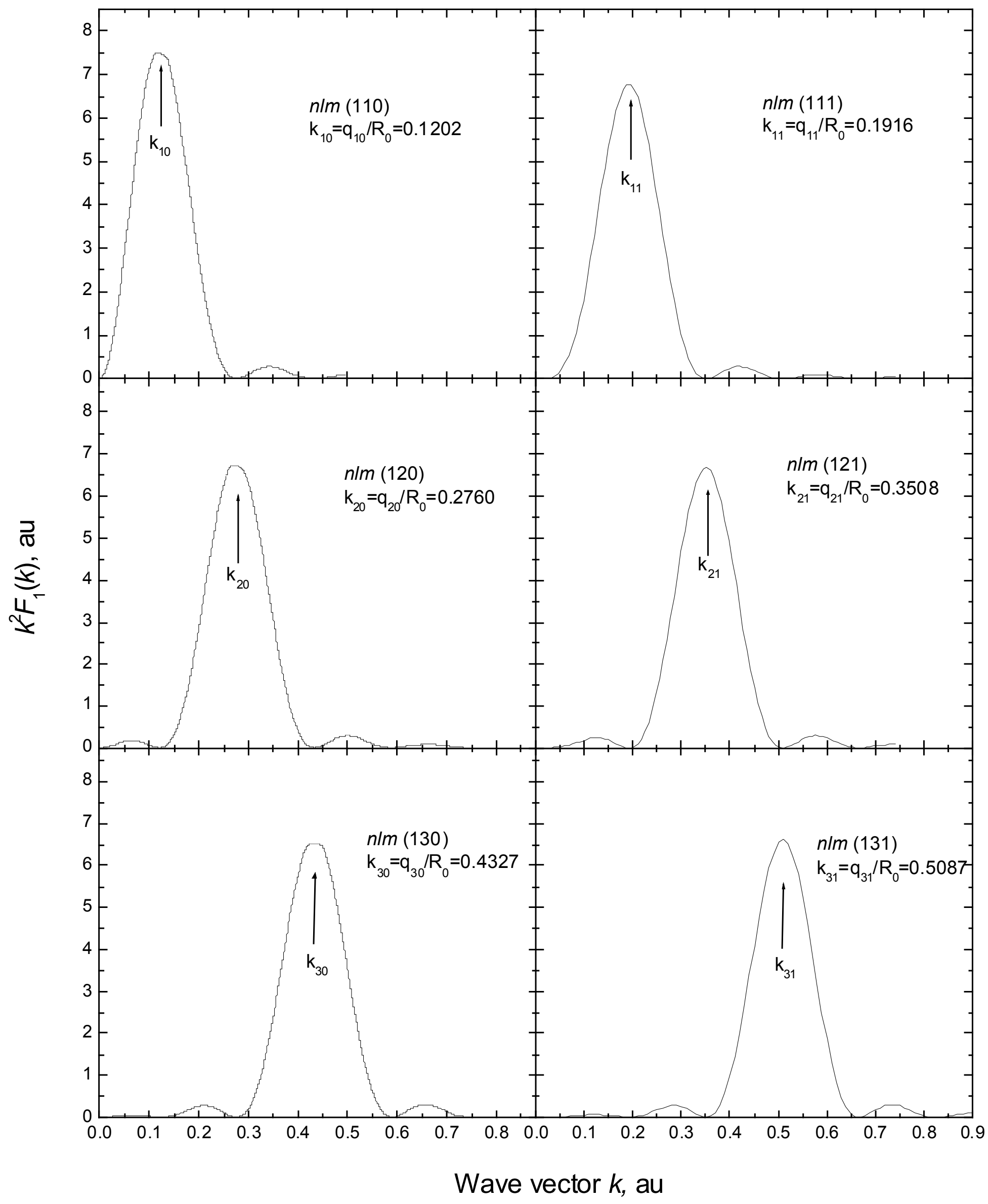

Fig. 4. The electron momentum *k*-distributions $k^2 F_1(k,\pi/2)$ for the lowest electronic states with the quantum numbers (*nlm*)

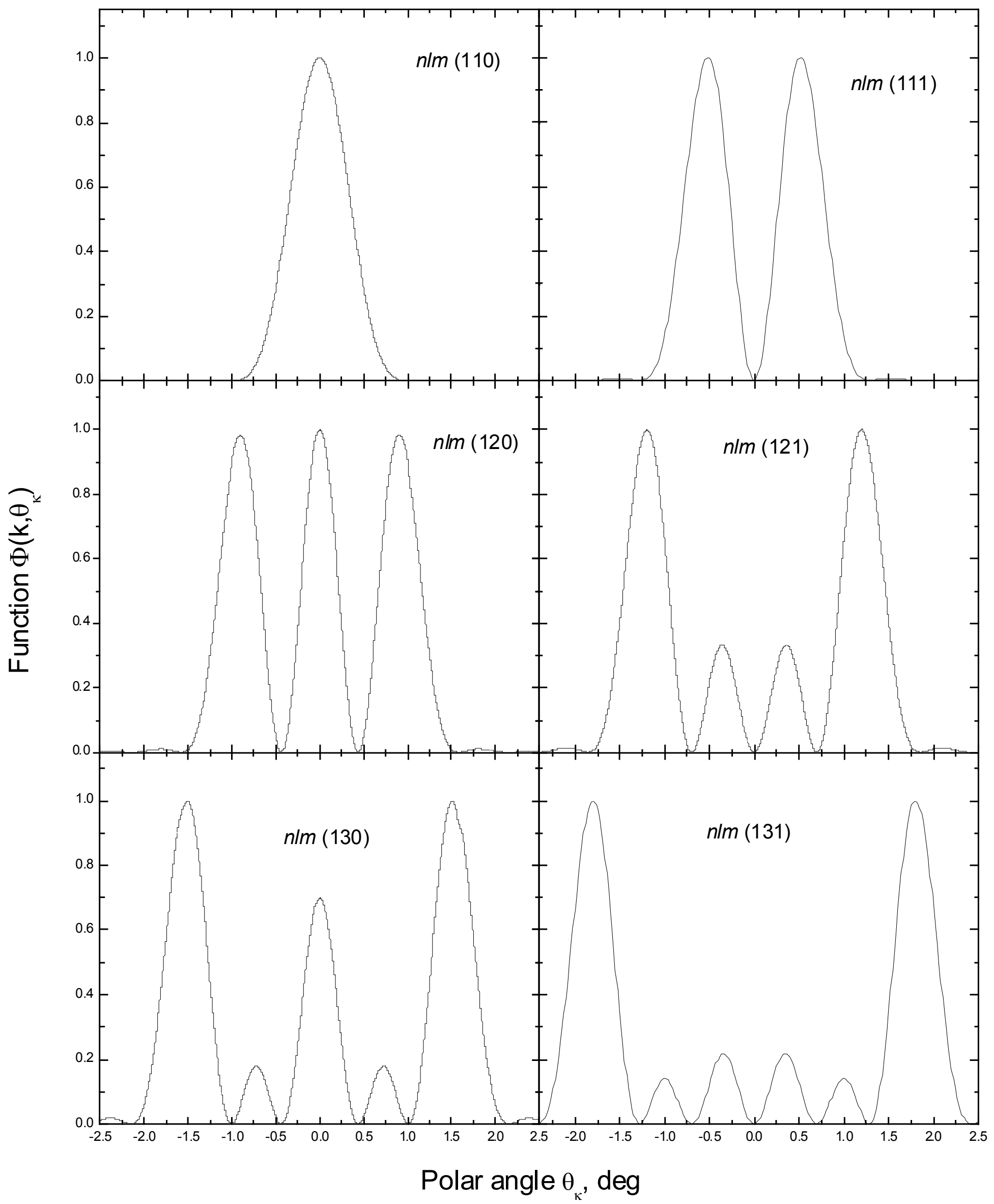

Fig. 5. The electron angular momentum distribution $\Phi(k,\vartheta_k)$ for some low-lying electronic states with the quantum numbers (*nlm*)

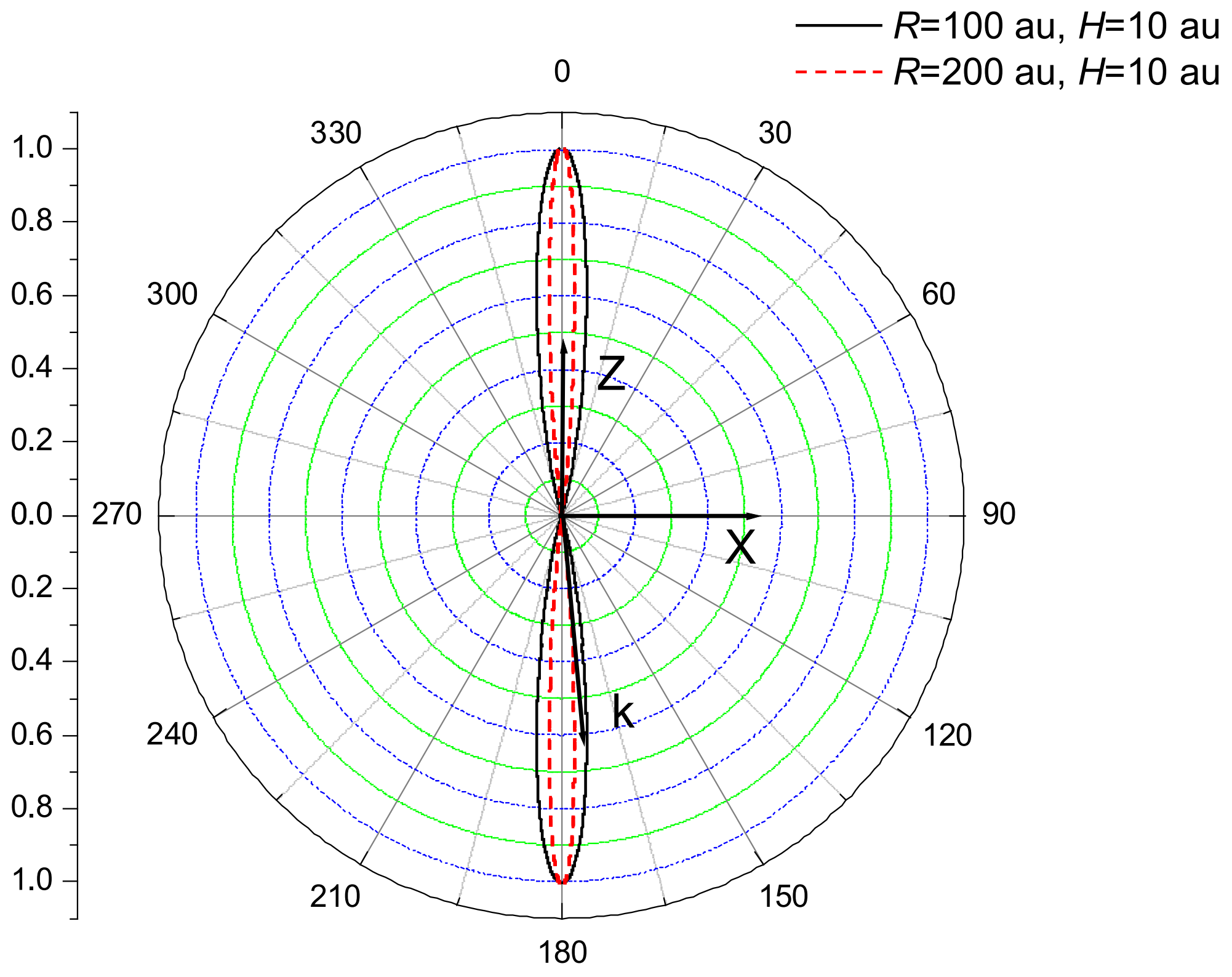

Fig. 6. The electron angular momentum distribution in the ground state (21) confined in the "short" cylindrical potential well (polar coordinates)

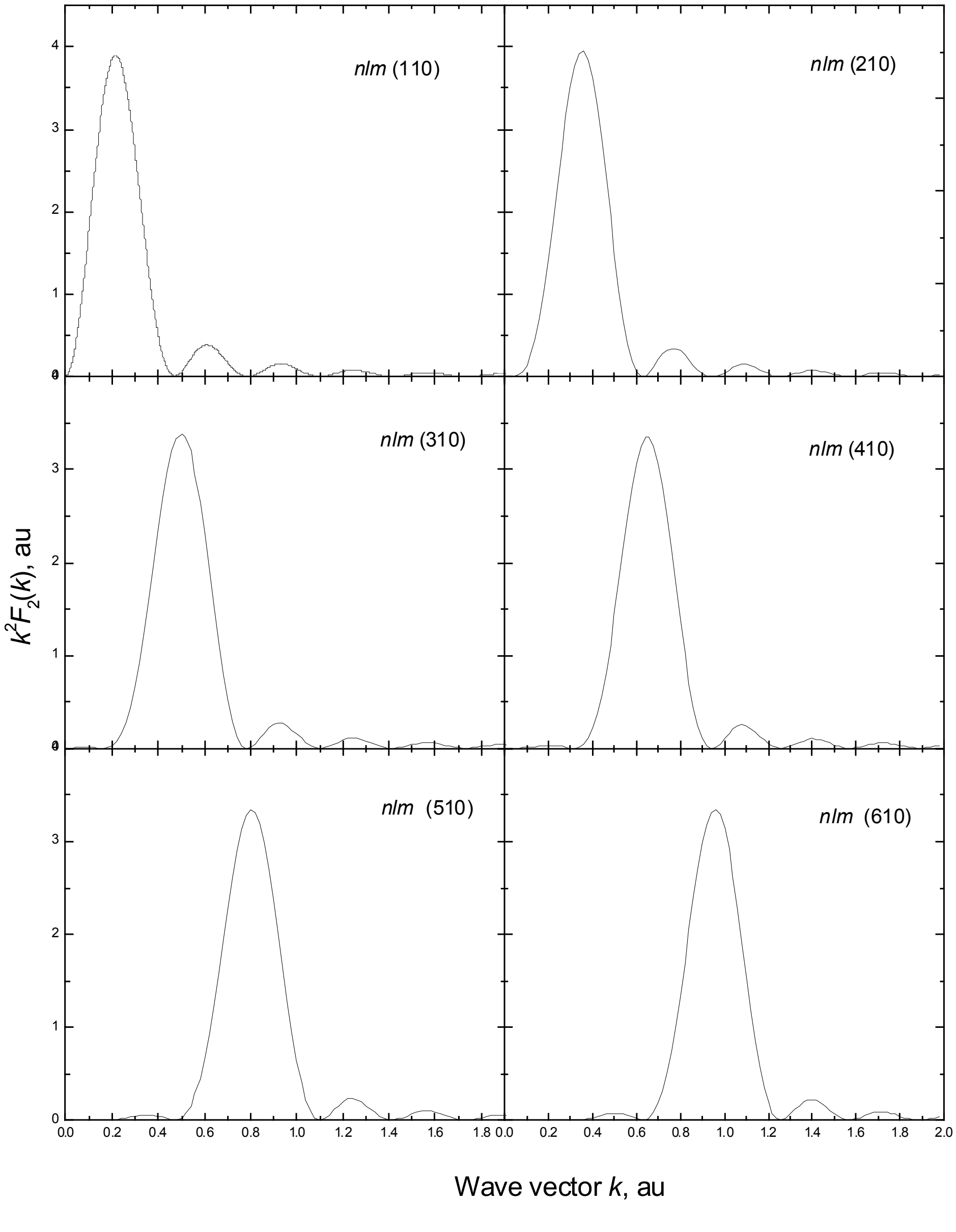

Fig.7. The electron momentum *k*-distributions $k^2 F_2(k,0)$ for the lowest electronic states with the quantum numbers (*nlm*)